\documentclass{eas}
\usepackage{graphicx}
\begin{document}

\newcommand{\GeVc}    {\mbox{$ {\mathrm{GeV}}/c                            $}}
\newcommand{\GeVcc}{\mbox{${\rm GeV/c^2}$}}
\newcommand{\MeVc}    {\mbox{$ {\mathrm{MeV}}/c                            $}}
\newcommand{\hetrois}    {\mbox{$ ^{3}{\mathrm{He}}                            $}}
\newcommand{\hetro}    {\mbox{$ ^{3}{\mathrm{He}}                            $}}
\newcommand{\xe}    {\mbox{$ ^{129}{\mathrm{Xe}}                            $}}
\newcommand{\ger}    {\mbox{$ ^{73}{\mathrm{Ge}}                            $}}
\newcommand{\al}    {\mbox{$ ^{27}{\mathrm{Al}}                            $}}
\newcommand{\fl}    {\mbox{$ ^{19}{\mathrm{F}}                            $}}
\newcommand{\tritium}    {\mbox{$ ^{3}{\mathrm{H}}                            $}}
\newcommand{\hequatre}    {\mbox{$ ^{4}{\mathrm{He}}                            $}}
\newcommand{\he}{$^4$He }
\newcommand{\hee}{$^4$He}
\newcommand{\het}{$^3$He }
\newcommand{\hett}{$^3$He}
\newcommand{\fe}{$^{55}$Fe }
\newcommand{\alu}{$^{27}$Al }
\newcommand{\cm}{$^{244}$Cm }
\newcommand{\iso}{C$_4$H$_{10}$ }
\newcommand{\isoo}{C$_4$H$_{10}$}
\newcommand{\nit}{N$_4$S$_3$ }
\newcommand{\neut}{$\tilde{\chi}^0$}
\newcommand{\neutt}{$\tilde{\chi}$}

\def\Journal#1#2#3#4{{#1} {\bf #2}, #3 (#4)}
\def\NCA{\em Nuovo Cimento}
\def\NIMA#1#2#3{{\rm Nucl.~Instr.~and~Meth.} {\bf{A#1}} (#2) #3}
\def\NIM#1#2#3{{\rm Nucl.~Instr.~and~Meth.} {\bf{#1}} (#2) #3}
\def\NPB{{\em Nucl. Phys.} B}
\def\PLB{{\em Phys. Lett.}  B}

\def\PRA#1#2#3{{\rm Phys. Rev.} {\bf{A#1}} (#2) #3}
\def\PRB#1#2#3{{\rm Phys. Rev.} {\bf{B#1}} (#2) #3}
\def\PRC#1#2#3{{\rm Phys. Rev.} {\bf{C#1}} (#2) #3}
\def\PRD#1#2#3{{\rm Phys. Rev.} {\bf{D#1}} (#2) #3}

\def\JHEP#1#2#3{{\rm JHEP} {\bf{#1}} (#2) #3}
\def\ZPC{{\em Z. Phys.} C}
\def\PRL#1#2#3{{\rm Phys.~Rev.~Lett.} {\bf{#1}} (#2) #3}
\def\PLB#1#2#3{{\rm Phys.~Lett.} {\bf{B#1}} (#2) #3}
\def\APP#1#2#3{{\rm Astropart.~Phys.} {\bf{B#1}} (#2) #3}
\def\APJ#1#2#3{{\rm Astrophys.~J.} {\bf{#1}} (#2) #3}
\def\APJS#1#2#3{{\rm Astrophys.~J.~Suppl.} {\bf{#1}} (#2) #3}
\def\AA#1#2#3{{\rm Astron. \& Astrophys.} {\bf{#1}} (#2) #3}
\def\JCAP#1#2#3{{\rm JCAP} {\bf{#1}} (#2) #3}


\title{MIMAC : a $\mu$TPC detector for non-baryonic dark matter search} 
\author{F. Mayet}\address{LPSC, Universit\'e Joseph Fourier Grenoble 1,
  CNRS/IN2P3, Institut Polytechnique de Grenoble, Grenoble, France}
\author{O. Guillaudin}\sameaddress{1}
\author{D. Santos}\sameaddress{1}
\author{A. Trichet}\sameaddress{1}

\begin{abstract}
The MIMAC project is multi-chamber detector for Dark Matter search, aiming at measuring both track and ionization with a matrix of micromegas $\mu$TPC filled with \hetrois~and $\rm CF_4$. 
Recent experimental results on the first measurements of the Helium quenching factor at low energy (1 keV recoil) are presented.
\end{abstract}
\runningtitle{F. Mayet {\it et al.} : $\mu$TPC detector for non-baryonic dark matter search}
\maketitle

\section{Introduction}
There is strong evidence in favor of a  dark  matter dominated Universe :  locally,  
from the rotation curves of spiral galaxies (\cite{rubin}) or
the Bullet cluster (\cite{clowe}) and on the largest scales, from  cosmological  observations (\cite{wmap,archeops,CG}). 
Most of the matter in the Universe consists of cold non-baryonic dark matter (CDM), the 
leading candidate for this class of yet undiscovered particles (WIMP)  being the 
lightest supersymmetric particle. 
In various supersymmetric scenarii (SUSY), this neutral and colorless particle is the lightest 
neutralino \neutt.\\
Tremendous experimental efforts on a host of techniques have been made in the field of direct search of
non-baryonic dark matter. Several detectors already reached sufficient sensitivity to begin to test regions of the 
SUSY parameter space. However, going  further required careful choice of  radiopurity of materials, underground 
laboratory, shielding, detector design  and 
data analysis (\cite{Moulin:2006pt}). As the expected event rates are very small (${\cal O} \rm (10^{-5}-1) \ day^{-1}kg^{-1}$), the challenge is to  
distinguish  a geniune WIMP signal from  backgrounds (mainly neutrons and $\gamma$-rays).
Two main dark matter search strategies may be identified : the detector may be designed to reach an  
extremely  high level of background rejection or to provide an unambigious positive signal.  
This can be achieved via :
\begin{itemize}
\item a dependence of the differential event rate on the atomic mass of 
the target nucleus, requiring a  good control of detector systematics,
\item a correlation of the signal with the motion of the detector with respect to the galactic 
Dark Matter halo. This can either be an annual modulation, due to the motion of the earth around the sun or  a strong  
direction dependence of the incoming neutralino, toward the Cygnus constellation to 
which points the sun's velocity vector (\cite{directionality,agreen}).
\end{itemize}
Several projects aiming at directionnal detection  of Dark Matter are being~developped (\cite{Drift,MIMAC,Collar,mit}).

\section{The MIMAC project}
The MIMAC project is multi-chamber detector for Dark Matter search. The idea is to measure both track and ionization with a matrix of micromegas $\mu$TPC filled with \hetrois~and $\rm CF_4$. The use of these two gases is motivated by their privileged features for dark matter search. With their odd atomic number, a detector made of such targets will be sensitive to the spin-dependent 
interaction, leading to a natural complementarity with existing detectors mainly sensitive to 
scalar interaction, in various SUSY models, e.g. non-universal SUSY (\cite{moulin,mayet-susy}).  
Moreover, as shown in (\cite{moulin}), a 10 kg \hetrois~dark matter detector  with a 1 keV threshold  (MIMAC) would present a sensitivity to SUSY models allowed by present cosmology and accelerator constraints. This study  highlights the complementarity of this experiment with most of current spin-dependent experiments : proton based detectors as well as $\nu$ telescopes.\\
Using both  \hetrois~and $\rm CF_4$ in a patchy matrix of $\mu$TPC opens the possibility to compare rates for two atomic masses, and to study neutralino interaction separately with neutrons and protons as the spin content of these nuclei is dominated by one kind of nucleon.\\
Low pressure operation of the MIMAC detector will enable the possibility to discriminate neutralino signal and background on the basis of track shape and incoming direction.

\section{Ionization quenching factor measurement}
As far as detection of Dark Matter is concerned, the ionization is one of the most important channels; heat, scintillation and tracking being the others. The ionization quenching factor (IQF) is defined as the fraction of energy released through ionization by a recoil in a medium  
compared with its kinetic energy. Measuring IQF, especially at low energies, is a key point for Dark Matter detectors, since it is needed to evaluate the nucleus recoil energy and hence the WIMP kinematics.\\
The energy released by a particle in a medium produces in an interrelated way three different processes: i) ionization, producing a number of electron - ion pairs, ii) scintillation, producing a number of photons through de-excitation of 
quasi-molecular states and iii) heat produced essentially by the motion of nuclei and electrons. 
The fraction of energy given to electrons has been estimated theoretically (\cite{Lindhard}) and parametrized by (\cite{Lewin}).\\ 
In the last decades an important effort  has been made to measure the IQF in different materials : 
gases (\cite{H}), solids (\cite{Ge,Si}) and liquids (\cite{Xe}), using different techniques.  The use of a monoenergetic neutron beam has been explored in solids with success (\cite{Qneutron1,Qneutron2}).
However in the low energy range the measurements are rare or absent for many targets (e.g. Helium) due to 
ionization threshold  of detectors and experimental constraints.\\
We have develop an experimental setup devoted to the measurement of low energy  (keV) ionization quenching factor.
The purpose is twofold : measuring for the first time the Helium quenching factor and performing very low energy measurements to test the prototype cell (resolution, threshold, ...) in the range of interest for Dark Matter (below 6 keV).\\

\begin{figure}[thb]
\begin{center}
\includegraphics[scale=0.4]{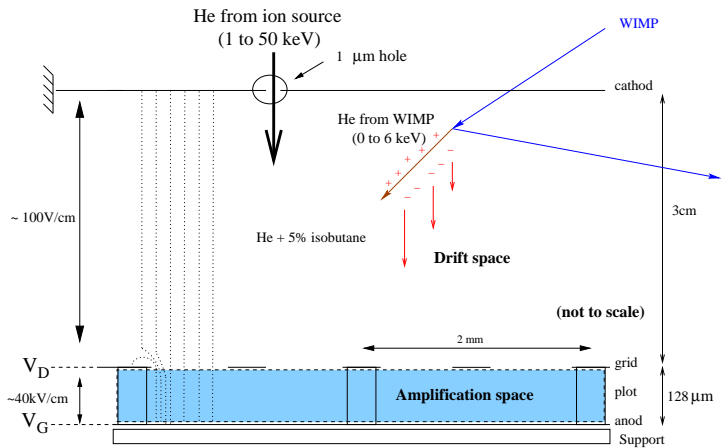}
\caption{Sketch of the Micromegas $\mu$TPC used for the Quenching factor measurement (not to scale).}
\label{fg.micromegas}
\end{center}
\end{figure}

The experimental set-up is the following  : an Electron
Cyclotron Resonance Ion Source (\cite{geller}) with
an extraction potential from a fraction of one kV up to 50
kV, is coupled to Micromegas (micromesh gaseous) detector via a $1 \ \mu m $ hole with a differential pumping.\\
Energy values have been previously checked by a time-of-flight measurements through a $\rm 50 \ nm$ thick $\rm N_4Si_3$ foil (\cite{Mayetnim}).
The ionization produced in the gas has been measured with a Micromegas (micromesh gaseous) detector~(\cite{Giomataris:1995fq}) 
adapted to a cathode integrated mechanically to the interface of the ion source. It is a  bulk type Micromegas (\cite{bulk}), 
in which the grid and the anode are built and integrated with a fixed gap, 128 $\mu m$ for  measurements between 
350 and 1300 mbar. The electric fields for the drift and the avalanche have been selected to optimize the transparence 
of the grid and the gain for each ion energy. Typical applied field were 
 $\rm 100 \ V/cm$ for the drift and a voltage of 450 V for the avalanche. The drift distance between the cathode and the grid was  3 cm, large enough 
to include the tracks of \he nuclei of energies up to 50 keV. These tracks, of the order of 6 mm for 50 keV, are roughly of the same length than the electrons tracks  produced by the X-rays emitted by th \fe source used for calibration.\\

In order to measure the quenching factor of \he in a gas mixture of 95\% of \he and 5\% of 
isobutane (\isoo) we proceed
as follows:  i) the ionization given by the Micromegas was calibrated by the two X-rays (1.486 and 5.97 keV) at each working point 
of the Micromegas defined by the drift voltage (V$_d$), the gain voltage (V$_g$) and the pressure, ii) the number of ions per 
seconde sent was kept lower than 25, 
to prevent any problem of recombination in primary charge 
collection or space charge effect.\\

Two different calibration sources have been used : the 1.486 keV X-rays of \alu produced by 
alpha particles emitted by a source of \cm under a thin foil of aluminium and a standard \fe X-ray source  giving the 5.9 keV K$_{\alpha}$    and the 6.4 keV K$_{\beta}$  lines.
These two lines, as they were not resolved by our detector, have been considered as a single one of  5.97 keV, taking into account their relative intensities. The IQF of a recoil will be the ratio between this energy and the  kinetic energy of such recoil. In such a way, the IQF compares the nuclei ionization efficiency  with respect to the electrons. The  ionization spectra of 1.5 keV \he nuclei  
and of electrons of roughly the same energy (1.486 keV) are shown on Figure \ref{plot2}. 

 \begin{figure}[thb]
\begin{center}
 \includegraphics[scale=0.38,angle=270]{mayet.fig2.epsi}
\caption{Spectra of 1.5 keV  kinetic energy \he (left) and 1.486 keV X-ray of
\alu (right) in \he +5\% \iso  mixture at 700 mbar.}  
\label{plot2}
\end{center}
\end{figure}

The measurement reported (\cite{prl}) have been focused on the  low energy \he IQF.  Figure \ref{plot2} presents the results at 700 mbar 
compared with the Lindhard theory for \he ions in pure \he and with respect to the SRIM 
simulation (\cite{srim})  for \he in the same gas mixture used during the measurements. We observe a difference between the SRIM simulation and the experimental
points of up to 20\% of the  kinetic energy of the
nuclei, shown in Fig. \ref{plot2}. This difference may be assigned to the scintillation produced by the \he nuclei in \he gas. 
This difference is reduced at lower pressures due to the fact that the amount of scintillation is reduced when the 
nucleai mean distance is increased giving a lower production probability of eximer states. 
More details may be found in (\cite{prl}).\\
As shown in (\cite{trichetlyon}) energy resolution of Micromegas $\mu$TPC has been measured down to 1 keV.
At 6 keV, it is of the order of 10\%. Moreover, it does not depend on pressure and it does not affect 
the number of expected events for Dark Matter, even at low energy. 
Threshold as low as 300 eV (ionization) has been reached.

As presented in (\cite{prl}), measurements have been done in various experimental conditions, showing 
a  clear increase at lower gaz pressure 
and also at lower isobutane percentage (the quencher). Hence, the ionization signal is expected to increase with Helium 
purity, which is needed for Dark Matter search, and also at the lowest pressures, which is compulsory for directionnal detection.\\

  \begin{figure}[htb]
\begin{center}
\includegraphics[scale=0.4,angle=270]{mayet.fig3.epsi}
\caption{
Helium ionization quenching factor  as a function of \he  kinetic energy (keV). Lindhard theory prediction 
($\rm ^4{He}$ in pure $\rm ^4{He}$ and $\rm ^3{He}$ in pure $\rm ^3{He}$), parametrized 
as in \cite{Lewin},  is presented and compared with SRIM simulation results (solid line and points) 
in the case of \he in \he+5\% \iso mixture. Measured quenching factor are presented at 700 mbar with error bars included 
mainly dominated by systematic errors. The differences between data and SRIM simulation are shown by triangles.}
\label{plot2}
\end{center}
\end{figure}

\section{Conclusion}
In summary, this first measurement of Helium ionization in Helium down 
to energies of 1 keV recoil opens the possibility to develop a Helium $\mu$TPC for Dark Matter.
For the next step, measurements of $\rm CF_4$ quenching and recoil tracks at low pressure are being investigated.
 

\end{document}